\documentclass[superscriptaddress,aps,letterpaper,10pt,twocolumn,floats,showpacs,amsmath,amsfonts,amssymb,pre,nofootinbib]{revtex4-2}
\usepackage{graphicx}
\usepackage[colorlinks=true,citecolor=blue]{hyperref}
\usepackage[caption=false]{subfig}
\usepackage{pstricks}
\usepackage{pst-node}
\usepackage{amsmath}
\usepackage{amsbsy}
\usepackage{verbatim}
\usepackage{dsfont}
\usepackage{MnSymbol}

\DeclareMathOperator*{\argmin}{arg\,min}

\begin{document}
	
	\title{Nonanalytic Landau functionals shaping the finite-size scaling of fluctuations and response functions in and out of equilibrium}
	
	\author{Krzysztof Ptaszy\'{n}ski}
 \email{krzysztof.ptaszynski@ifmpan.poznan.pl}
\affiliation{Complex Systems and Statistical Mechanics, Department of Physics and Materials Science, University of Luxembourg, 30 Avenue des Hauts-Fourneaux, L-4362 Esch-sur-Alzette, Luxembourg}
	\affiliation{Institute of Molecular Physics, Polish Academy of Sciences, Mariana Smoluchowskiego 17, 60-179 Pozna\'{n}, Poland}
	
	\author{Massimiliano Esposito}
	\affiliation{Complex Systems and Statistical Mechanics, Department of Physics and Materials Science, University of Luxembourg, 30 Avenue des Hauts-Fourneaux, L-4362 Esch-sur-Alzette, Luxembourg}
	
	\date{\today}
	
	\begin{abstract}
 Landau theory relates phase transitions to the minimization of the Landau functional (e.g., free energy functional), which is expressed as a power series of the order parameter. It has been shown that the critical behavior of certain physical systems can be described using Landau functionals that include nonanalytic terms, corresponding to odd or even noninteger powers of the absolute value of the order parameter. In particular, these nonanalytic terms can determine the order of the phase transition and the values of the critical exponents. Here, we show that such terms can also shape the finite-size scaling behavior of fluctuations of observables (e.g., of energy or magnetization) or the response functions (e.g., heat capacity or magnetic susceptibility) at the continuous phase transition point. We demonstrate this on two examples, the equilibrium molecular zipper and the nonequilibrium version of the Curie--Weiss model.
	\end{abstract}
	
	\maketitle
	
	\section{Introduction}
Landau theory of phase transitions~\cite{landau1965collected} is one of the cornerstones of modern statistical mechanics and condensed matter physics. It explains the mechanism of equilibrium phase transitions by postulating the minimization of the free energy functional, also referred to as the \textit{Landau functional}. It assumes that this functional can be expanded as a power series of a macroscopic order parameter $\phi$ (e.g., magnetization for magnetic systems). Depending on the presence of the cubic term ($\propto \phi^3) $ and the sign of the quartic term ($\propto \phi^4$) of the expansion, the phase transition may be either continuous or discontinuous. For continuous phase transitions, the theory predicts universal scaling of the order parameter close to the critical temperature $T_c$. It is described by a power law $\phi \propto (1-T/T_c)^{\hat{\beta}_\text{MF}}$, with the mean field critical exponent $\hat{\beta}_\text{MF}=1/2$. While the quantitative applicability of the Landau theory (including the predicted values of the critical exponents) is confined to mean field models, it gave rise to a more advanced Landau--Ginzburg--Wilson theory~\cite{wilson1974renormalization} that takes into account fluctuations of the order parameter, which are relevant for finite-dimensional systems. The concept of Landau functional has been further generalized to nonequilibrium systems~\cite{aron2020landau} using methods of large deviation theory~\cite{Touchette2009, FalascoReview}.

Standard Landau theory assumes that the Landau functional is an analytic function of the order parameter. However, it was shown that certain physical systems can be described using Landau functionals containing nonanalytic terms, e.g., related to odd, and even noninteger powers of the absolute value of the order parameter, or expressed as its logarithmic function. This has been observed for thermal and quantum phase transitions in the presence of soft modes (e.g., in liquid crystals or quantum ferromagnets)~\cite{belitz2005how} and
nonequilibrium Ising models~\cite{aron2020landau, aron2021nonanalytic}. Similar nonanalytic terms have also been encountered in the theory of periodically sheared soft matter~\cite{mari2022absorbing}. In this paper, we add another example, namely, a generalization of the molecular zipper model~\cite{kittel1969phase}. Significantly, the nonanalytic terms of the Landau functional can determine the critical behavior of the system, including the order of the phase transition and the values of the critical exponents.

Genuine phase transitions, associated with the nonanalytic behavior of the order parameter, strictly speaking, occur only in the thermodynamic limit of the infinite system size. As a consequence, critical exponents may be difficult to determine directly using simulations and experiments on finite systems. This problem is often dealt with employing finite-size scaling theory~\cite{ferdinand1969bounded, fisher1972scaling,botet1983large} to extract the values of critical exponents from the finite-size scaling behavior of observables and response functions~\cite{landau1976finite, binder1981critical, landau2021guide}. Here, we show that in systems with nonanalytic Landau functionals, the nonanalytic terms can determine the finite-size scaling of fluctuations of the system observables (e.g., of its energy or magnetization) and the response functions (e.g., the heat capacity or magnetic susceptibility) at the phase transition point. This provides a way to detect their presence by analyzing finite-size systems. We demonstrate this on two examples: a generalization of the molecular zipper proposed by Kittel~\cite{kittel1969phase} as a toy model of the unwinding transition in DNA, and the nonequilibrium Curie--Weiss model coupled to two baths with different spectral densities, which has previously been used to illustrate the concept of nonequilibrium Landau functionals with nonanalytic terms~\cite{aron2020landau}. 

The paper is organized as follows: in Secs.~\ref{sec:zipp} and~\ref{sec:cw} we present the models considered and discuss their finite-size scaling behavior, while in Sec.~\ref{sec:concl} we present the conclusions that follow from our results.
	
	\section{Molecular zipper} \label{sec:zipp}
 \subsection{Equilibrium thermodynamics}
 %
\begin{figure}
	\centering
	\includegraphics[width=0.9\linewidth]{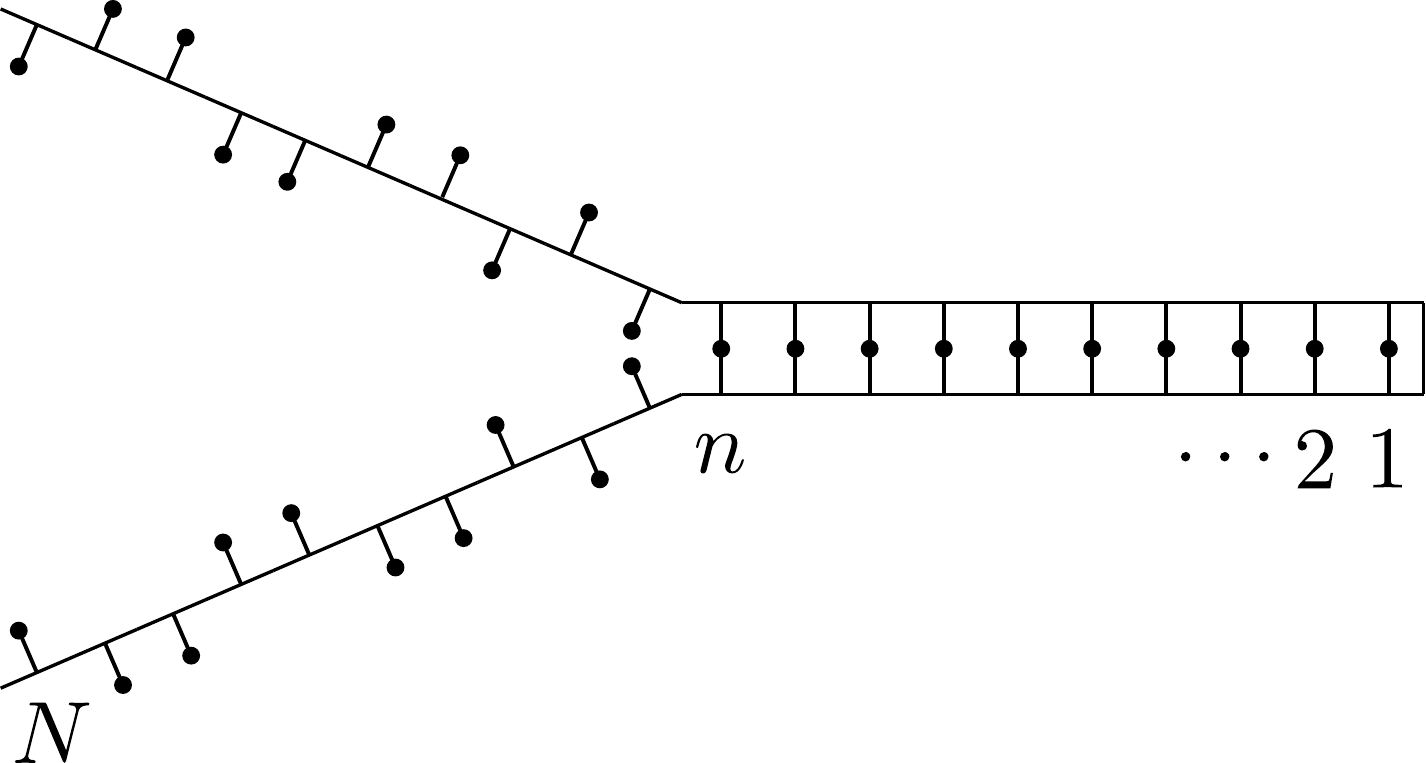}
	\caption{Scheme of the molecular zipper with $n=10$ closed links and $N-n=10$ open links. The scheme corresponds to the case of $g=2$, with open links taking two orientations: ``inward'' and ``outward''.}
	\label{fig:zipperscheme}
\end{figure}
%
We first consider a generalization of the molecular zipper model proposed by Kittel~\cite{kittel1969phase} to qualitatively describe the DNA denaturation. It later received interest in other contexts, such as large deviation theory~\cite{touchette2010, deger2018lee}, nonequilibrium dynamics and thermodynamics~\cite{holubec2012dynamics}, or melting of thin films~\cite{abdullah2015zipper}. In our discussion, we do not aim to describe the behavior of any particular real-world system, but rather treat the molecular zipper as a toy model of an equilibrium phase transition. This might provide insight into the behavior of more complex equilibrium systems with nonanalytic Landau functionals, such as the systems with soft modes~\cite{belitz2005how}. We also notice that many elements of the model description are shared with our previous work focusing on its dynamical properties~\cite{ptaszynski2024finite} (however, the case analyzed there exhibited a discontinuous rather than a continuous phase transition). We reproduce them here to make the paper self-contained.

The model consists of a double-stranded macromolecule, rigidly connected at one end, stabilized by $N$ parallel links that can be either closed or open (Fig.~\ref{fig:zipperscheme}). The $i$th link can close only if the $i-1$ preceding links are also closed. Closing the $i$th link decreases the energy of the system by $\epsilon_i$. (This generalizes the original model, where all energies $\epsilon_i$ were equal to each other). The energy of the system with $n$ closed links is equal to
\begin{align} \label{eq:enzippdisc}
E_n=-\sum_{i=1}^n \epsilon_i \,.
\end{align}
It is also assumed that the link can be opened in $g$ different energy-degenerate ways (e.g., open links can be oriented in different directions). The system with $n$ closed links corresponds then to $g^{(N-n)}$ different microscopic configurations of the system. The Boltzmann entropy of the system with $n$ closed links is thus equal to 
\begin{align} \label{eq:entrzippdisc}
S_n=k_B \ln g^{(N-n)}=(N-n) k_B \ln g \,.
\end{align}

We further define the free energy functional of the system $F_n=E_n-T S_n$, where $T$ is the temperature. The probability that $n$ links are closed is given by the Boltzmann distribution
\begin{align} \label{eq:probn}
p_n = \frac{e^{-\beta F_n}}{\sum_{m=0}^N e^{-\beta F_m}} \,.
\end{align}

To ensure that the equilibrium free energy of the model is extensive with system size, we take the energies of the closed links $\epsilon_i$ to be parameterized by a scale-invariant function $f(x)$,
\begin{align}
	\epsilon_i=f\left (\frac{i}{N} \right) \,.
\end{align}
We then consider the limit of the large system size $N$ and parameterize $n$ with the rescaled variable $q=n/N \in[0,1]$. We define the free energy density functional
\begin{align} \label{eq:freedenzippdef}
\mathcal{F}(q) \equiv \lim_{N \rightarrow \infty} \frac{F_n}{N} \quad \text{for} \quad q=\frac{n}{N} \,.
\end{align}
Taking a continuous limit of Eqs.~\eqref{eq:enzippdisc}--\eqref{eq:entrzippdisc}, the free energy density functional can be expressed as
\begin{align} \label{eq:freeendenzippgen}
\mathcal{F}(q) =-\int_0^q f(x) dx-(1-q) k_B T \ln g \,.
\end{align}
Comparing Eqs.~\eqref{eq:probn} and~\eqref{eq:freedenzippdef}, we observe that the probability that $n$ links are closed exhibits the large deviation property
\begin{align} \label{eq:probnldev}
p_n \propto e^{-N \beta \mathcal{F}(n/N)} \,,
\end{align}
where $\beta=1/(k_B T)$. For large $N$, the probability distribution becomes narrowly focused around the minimum of $\mathcal{F}(q)$. Thus, the average number of closed links scales asymptotically as
\begin{align}
\lim_{N \rightarrow \infty} \langle n \rangle_\text{eq}/N = q_\text{eq} \,,
\end{align}
where
\begin{align} \label{eq:qeq}
q_\text{eq}=\operatorname*{arg\,min}_{q \in [0,1]} \mathcal{F}(q)
\end{align}
is the value of $q$ minimizing $\mathcal{F}(q)$. Minimization is restricted here to the domain of $q$, that is, the interval $[0,1]$.

We further focus on a particular model where the energies of closed links are given by the function
\begin{align}
	f(x)=\epsilon(1-x^\alpha) \,,
\end{align}
where $\alpha >0$. Using Eq.~\eqref{eq:freeendenzippgen}, the free energy density functional can be expressed as
\begin{align} \label{eq:freeendenzippspec}
	\mathcal{F}(q) =(k_B T \ln g-\epsilon)q+\frac{\epsilon q^{\alpha+1}}{\alpha+1} -k_B T \ln g \,.
\end{align}
Notably, its second term becomes nonanalytic at $q=0$ when $\alpha$ is a noninteger number.

 %
\begin{figure}
	\centering
	\includegraphics[width=0.9\linewidth]{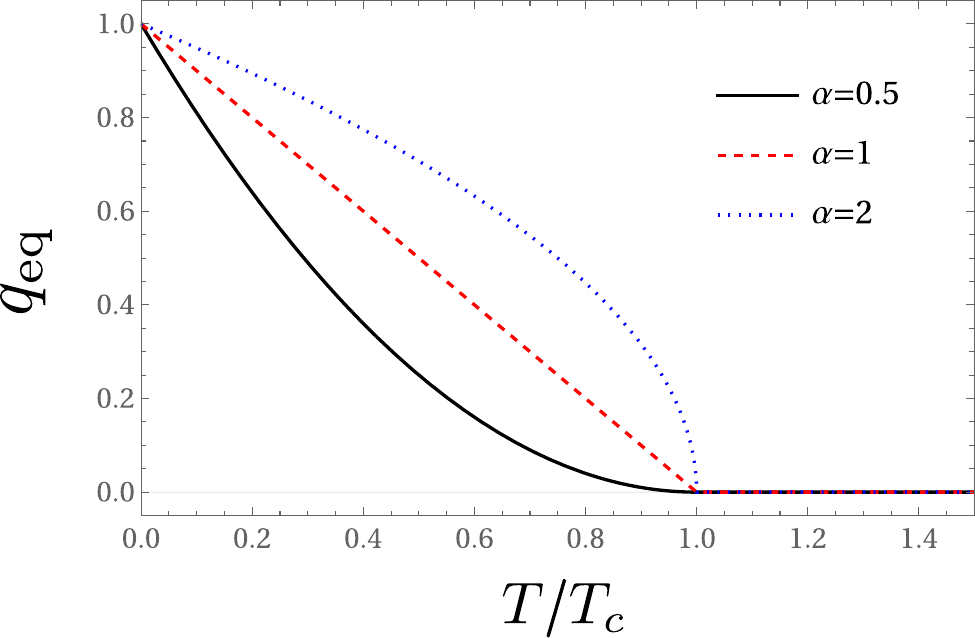}
	\caption{Temperature-dependence of the order parameter $q_\text{eq}$ for different values of the parameter $\alpha$.}
	\label{fig:orderpar}
\end{figure}
%
Minimizing $\mathcal{F}(q)$ over $q$, we find that the system exhibits a continuous phase transition at the critical temperature
\begin{align}
T_c=\frac{\epsilon}{k_B \ln g} \,,
\end{align}
with the order parameter $q_\text{eq}$ behaving as
\begin{align}
	q_\text{eq}=\begin{cases}
		0 \quad &\text{for} \quad T \geq T_c \,. \\
		\left(1-\frac{T}{T_c} \right)^{1/\alpha} \quad &\text{for} \quad T<T_c \,.
	\end{cases}
\end{align}
Thus, the parameter $\alpha$ determines the critical exponent of the phase transition $1/\alpha$ (which here characterizes the scaling of $q_\text{eq}$ for all $T<T_c$, not only close to $T_c$). The temperature dependence of $q_\text{eq}$ is plotted, for a few values of $\alpha$, in Fig.~\ref{fig:orderpar}. We note that the observed behavior contrasts with the previously considered version of the model~\cite{kittel1969phase,touchette2010,abdullah2015zipper,holubec2012dynamics,ptaszynski2024finite,deger2018lee} where the phase transition was discontinuous. This is due to the chosen form of the energy function $f(x)$, which here decreases monotonically with $x$, while in the cases studied previously it was either constant or monotonically increasing.

Let us here note that, in the model considered, the parameter $q$ admits only positive values. Thus, the nature of the studied phase transition differs from the commonly known symmetry-breaking phase transitions where, at the phase transition point, a single minimum of the free energy density functional splits into two degenerate ones, corresponding to a positive and negative value of the order parameter. Such transitions can only occur when the Landau functional is an even function of the order parameter, which is not the case here. In fact, the symmetry of the system considered is always broken, since the molecule is rigidly connected at one end.

Finally, we analyze the order of the phase transition according to the Ehrenfest classification. Within this framework, the transition is of the $j$th order if the $j$th temperature derivative of the free energy is the lowest, which is discontinuous at the critical temperature $T_c$. Since the equilibrium free energy scales in the thermodynamic limit as $\lim_{N \rightarrow \infty} F_\text{eq}/N = \mathcal{F}(q_\text{eq})$, we get
\begin{align}
\lim_{N \rightarrow \infty} \frac{F_\text{eq}}{N} = \begin{cases}
		-k_B T \ln g \; &\text{for} \; T \geq T_c \,, \\
		-\frac{\epsilon \alpha}{\alpha+1} \left(1-\frac{T}{T_c} \right)^{\frac{\alpha+1}{\alpha}} -k_B T \ln g \; &\text{for} \; T<T_c \,.
	\end{cases}
\end{align}
We find that the phase transition is of the $j$th order if
\begin{align} \label{eq:ordzipp}
    \frac{1}{j-1} \leq \alpha <\frac{1}{j-2} \,.
\end{align}
We recall that in our model the phase transition is continuous, so that $j \geq 2$. In particular, for $\alpha=1/(j-1)$, the $j$th derivative of the free energy is discontinuous but finite at $T=T_c$. Otherwise, it is divergent. Thus, by tailoring the exponent $\alpha$, one can modify the order of the phase transition (in principle, to an arbitrary value).

\subsection{Fluctuations of the number of closed links}
We now turn to the main topic of the paper, namely, the finite-size scaling of fluctuations at the phase transition point (i.e., for $T=T_c$). First, we consider the variance of the number of closed links,
\begin{align}
\langle \Delta n^2 \rangle = \langle n^2 \rangle-\langle n \rangle^2 \,,
\end{align}
where the moments of $n$ are calculated as
\begin{align}
\langle n^k \rangle= \sum_{n=0}^N p_n n^k \,,
\end{align}
with $p_n$ given by Eq.~\eqref{eq:probn}. In the continuous limit, the moments of $n$ can be approximated using Eq.~\eqref{eq:probnldev} as
\begin{align}
\langle n^k \rangle \approx N^k \frac{\int_0^\infty q^k e^{-\beta N \mathcal{F}(q)} dq}{\int_0^\infty e^{-\beta N \mathcal{F}(q)} dq} \,.
\end{align}
Here we extend the integration range up to $\infty$, since for large $N$ the exponential function $\exp[-\beta N\mathcal{F}(q)]$ decays very quickly with $q$. We note that, at $T=T_c$, the term of $\mathcal{F}(q)$ that is linear in $q$ vanishes, and only the term proportional to $q^{\alpha+1}$ remains. The integral yields the approximate formula
\begin{figure}
	\centering
	\includegraphics[width=0.9\linewidth]{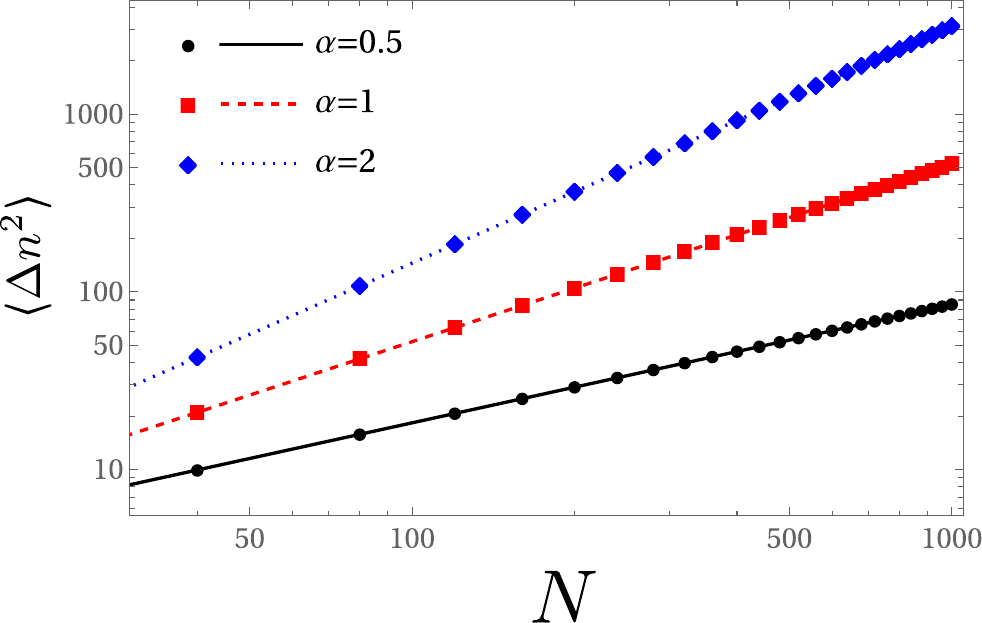}
	\caption{The variance of the number of closed links at the critical point $T=T_c$ as a function of the system size $N$, plotted on the log-log scale, calculated exactly (dots) and using the approximate formula~\eqref{eq:varlinksapp} (lines).}
	\label{fig:numberlinks}
\end{figure}
\begin{align} \label{eq:varlinksapp}
\langle \Delta n^2 \rangle \approx N^{\frac{2 \alpha }{\alpha +1}} \frac{ \left(\frac{\alpha +1}{\ln 2}\right)^{\frac{2}{\alpha +1}} \left[\Gamma \left(\frac{1}{\alpha +1}\right) \Gamma \left(\frac{3}{\alpha +1}\right)-\Gamma \left(\frac{2}{\alpha +1}\right)^2\right]}{\Gamma \left(\frac{1}{\alpha +1}\right)^2} \,,
\end{align}
where $\Gamma$ is the gamma function. This formula implies that the variance of the number of closed links follows the power-law scaling with the system size, with the exponent determined by the parameter $\alpha$. In Fig.~\ref{fig:numberlinks} we compare this approximate formula (lines) with the exact finite-size results (dots). We can observe a perfect agreement.

\subsection{Energy fluctuations and heat capacity}
Let us now consider the variance of the system energy
\begin{align}
\langle \Delta E^2 \rangle = \langle E^2 \rangle-\langle E \rangle^2 \,,
\end{align}
where the moments of energy read
\begin{align}
\langle E^k \rangle= \sum_{n=0}^N p_n E_n^k \,.
\end{align}
We further note that the energy variance is related to the heat capacity via the formula
\begin{align}
\langle \Delta E^2 \rangle = k_B T^2 \frac{\partial \langle E \rangle}{\partial T} \,.
\end{align}

Taking the continuous limit, the moments of energy can be approximated as
\begin{align}
\langle E^k \rangle \approx N^k \frac{\int_0^\infty dq \mathcal{E}(q)^k e^{-\beta N \mathcal{F}(q)}}{\int_0^\infty dq e^{-\beta N\mathcal{F}(q)}} \,,
\end{align}
where 
\begin{align}
\mathcal{E}(q) \equiv \lim_{N \rightarrow \infty} \frac{E_n}{N} \quad \text{for} \quad q=\frac{n}{N} \,
\end{align}
is the energy density functional which can be calculated as
\begin{align} 
\mathcal{E}(q) =-\int_0^q f(x) dx=-\epsilon \left( q-\frac{q^{\alpha+1}}{\alpha+1} \right) \,.
\end{align}
This yields an approximate expression for the energy variance,
\begin{align} \label{eq:varenapp}
\langle \Delta E^2 \rangle \approx A_1 N^{\frac{2\alpha}{\alpha+1}} +A_2 N^{\frac{\alpha}{\alpha+1}}+A_3 \,,
\end{align}
 %
\begin{figure}
	\centering
	\includegraphics[width=0.9\linewidth]{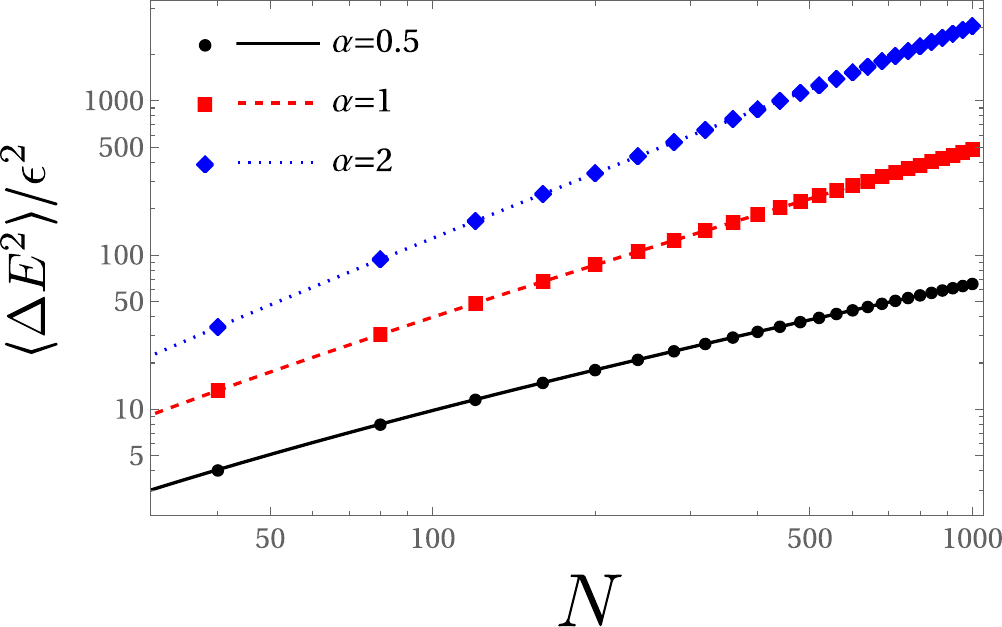}
	\caption{The energy variance at the critical point $T=T_c$ as a function of the system size $N$, plotted on the log-log scale, calculated exactly (dots) and using the approximate formula~\eqref{eq:varenapp} (lines).}
	\label{fig:energyfluct}
\end{figure}
%
with
\begin{align} \nonumber
A_1 &= \frac{ \epsilon^2 \left(\frac{\alpha +1}{\ln 2}\right)^{\frac{2}{\alpha +1}} \left[\Gamma \left(\frac{1}{\alpha +1}\right) \Gamma \left(\frac{3}{\alpha +1}\right)-\Gamma \left(\frac{2}{\alpha +1}\right)^2\right]}{\Gamma \left(\frac{1}{\alpha +1}\right)^2} \,, \\
A_2 &=-\frac{\epsilon^2 \left(4 \alpha +4\right)^{\frac{1}{\alpha +1}} (\ln 2) ^{-\frac{\alpha +2}{\alpha +1}} \Gamma \left(\frac{1}{2}+\frac{1}{\alpha +1}\right)}{\sqrt{\pi } (\alpha +1)} \,, \\ \nonumber
A_3 &= \frac{\epsilon^2}{(\alpha +1) (\ln 2)^2} \,.
\end{align}
It appears that the second and third terms in Eq.~\eqref{eq:varenapp} are still important for $N$ of the order of $10^2$.  In Fig.~\ref{fig:energyfluct} we compare the obtained approximate formula for the energy variance (lines) with the exact finite-size results (dots). We can again observe a perfect agreement. 

 \section{Nonequilibrium Curie-Weiss model} \label{sec:cw}
 \subsection{Model}
 We now consider a nonequilibrium open system with an effective nonanalytic Landau functional, which was originally proposed in~ Ref.~\cite{aron2020landau}. It is based on the Curie--Weiss model, a paradigmatic model of the paramagnetic--ferromagnetic transition~\cite{kochmanski2013curie}. It consists of $N$ spins that interact via the all-to-all Ising interaction. The energy of a particular spin configuration can be written as
\begin{align}
E=-\frac{J}{2N} \sum_{i,j=1}^N  \sigma_i \sigma_j-h \sum_{i=1}^N \sigma_i \,,
\end{align}
where $J \geq 0$ is the ferromagnetic Ising interaction, and $h$ is the magnetic field. Spins $\sigma_i$ are here the classical random variables with values $\pm 1$. The total magnetization of the system can be defined as $M=\sum_i \sigma_i \in \{-N,-N+2,\ldots,N\}$. Due to the all-to-all nature of the coupling, the energy of the system can be written in terms of the total magnetization,
\begin{align} \label{eq:encwm}
	E_M=-\frac{J}{2N} M^2-hM \,.
\end{align}
The system interacts with two thermal baths $i \in \{1,2\}$ with the temperatures $T_1$ and $T_2$. It is also assumed that each bath is coupled with an equal strength to each spin, and that they induce a Markovian flipping of individual spins. Then, the dynamics of the system can be described by a classical master equation for probabilities $p_M$ that the system has magnetization $M$~\cite{herpich2020njp},
\begin{align}
\dot{p}_M=\sum_{\pm} \left( W_{M,M\pm 2} p_{M \pm 2} - W_{M \pm 2,M} p_M \right) \,,
\end{align}
where $W_{M \pm 2,M}$ is the transition rate from the state with magnetization $M$ to the state with magnetization $M \pm 2$. The equation can be rewritten in the matrix form
\begin{align} \label{eq:masteqmat}
\dot{\boldsymbol{p}}=\mathbb{W} \boldsymbol{p} \,,
\end{align}
where $\boldsymbol{p}=(p_{-N},p_{-N+2},\ldots,p_N)^T$ is the vector of state probabilities, and $\mathbb{W}$ is the rate matrix with the off-diagonal elements $W_{kl}$ and the diagonal elements $W_{kk}=-\sum_{l \neq k} W_{lk}$. The stationary state of the system $\boldsymbol{p}^\text{st}$ is given by the condition
\begin{align}
\mathbb{W} \boldsymbol{p}^\text{st}=0 \,.
\end{align}

The transition rates can be further decomposed as a sum of contributions associated with each bath: $W_{M \pm 2,M}=W_{M\pm 2,M}^1+W_{M \pm 2,M}^2$. The individual contributions can be written as
\begin{align}
W_{M \pm 2,M}^i=\frac{N \mp M}{2N} C_i\left(\beta_i,E_{M \pm 2}-E_M \right) \,,
\end{align}
where $C_i(\beta_i,\omega)$ is the correlation function of the bath. Following Ref.~\cite{aron2020landau}, we use a model of correlations functions with a power-law spectral density of the bath,
\begin{align} \label{eq:specdens}
C_i(\beta_i,\omega)=\begin{cases}
		\gamma_i |\omega/2|^{\alpha_i} n(\beta_i \omega) &\text{for} \quad \omega \geq 0\,, \\
		 \gamma_i |\omega/2|^{\alpha_i} \left[1+n(-\beta_i \omega) \right] \quad &\text{for} \quad \omega<0 \,,
	\end{cases}
\end{align}
where $n(x)=[\exp(x)-1]^{-1}$ is the Bose-Einstein distribution. This corresponds to the excitation (relaxation) of the system for $\omega>0$ ($\omega<0)$, induced by the bosonic bath with the spectral density $\gamma_i |\omega/2|^{\alpha_i}$. Such models with power-law spectral densities of the bath (called Ohmic for $\alpha_i=1$, sub-Ohmic for $\alpha_i<1$, and super-Ohmic for $\alpha_i>1$) are commonly considered in the theory of open quantum systems~\cite{nazarov2009quantum,weiss2012quantum}. In particular, Ref.~\cite{aron2020landau} proposed to realize different exponents $\alpha_i$ by using $d_i$-dimensional bosonic baths with dispersion relations $\omega \propto k^{z_i}$, $k$ being the wavevector, so that $\alpha_i=d_i/z_i$. We also emphasize that to observe the phenomena described later, it is enough that the power-law scaling of the spectral density given by Eq.~\eqref{eq:specdens} occurs in the low-frequency range, rather than for all frequencies $\omega$~\cite{aron2020landau}.

\subsection{Nonequilibrium Landau functional}
Since the model is out of equilibrium, its state cannot be determined by minimizing the free energy functional. Nevertheless, the stationary state probabilities obey the large deviation principle~\cite{Touchette2009, FalascoReview} similar to Eq.~\eqref{eq:probnldev},
\begin{align} \label{eq:probmldev}
p_M^\text{st} \propto e^{-N V(M/N)} \,,
\end{align}
where $V(m)$ is the nonequilibrium quasipotential (also called rate function), which is a function of the rescaled magnetization $m=M/N$. This quasipotential plays the role of the Landau functional. For the model considered, it can be calculated as~\cite{ptaszynski2024critical}
\begin{align} \label{eq:quasipotform}
V(m)=\frac{1}{2} \int_{-1}^m dq \ln \frac{w_-(q)}{w_+(q)} \,,
\end{align}
with the scaled transition rates $w_\pm(m)$ defined as
\begin{align}
w_\pm(m) \equiv \lim_{N \rightarrow \infty} \frac{W_{M \pm 2,M}}{N}= \frac{(1 \mp m)}{2} \sum_{i=1}^2C_i[\beta_i,\mp 2 (Jm+h)] \,.
\end{align}
Equation~\eqref{eq:quasipotform} can be derived by using the detailed balance condition $W_{M,M+2}p^\text{st}_{M+2}  =W_{M+2,M} p^\text{st}_M $ to express the stationary probabilities as~\cite{landauer1962fluctuations, hanggi1982stochastic, vellela2009stochastic, ptaszynski2024critical}
\begin{align} \label{eq:stprob} \nonumber
&p_M^\text{st} = p_{-N}^\text{st} \frac{W_{-N+2,-N} W_{-N+4,-N+2} \ldots W_{M,M-2}}{W_{-N,-N+2} W_{-N+2,-N+4} \ldots W_{M-2,M}} \\ 
&=p_{-N}^\text{st} \exp \left(\ln \frac{W_{-N+2,-N}}{W_{-N,-N+2}} + \ldots + \ln\frac{ W_{M,M-2}}{ W_{M-2,M}} \right) \,.
\end{align}
Taking the limit $N \rightarrow \infty$, using the relation between a limit of the Riemann sum and a definite integral, and noting Eq.~\eqref{eq:probmldev}, we get Eq.~\eqref{eq:quasipotform}.

Let us now take $\alpha_2>\alpha_1$ and denote $\nu=\alpha_2-\alpha_1$. We further focus on the case of $h=0$ (the generalization to finite $h$ will be considered in Sec.~\ref{subsec:susc}). Then, the quasipotential $V(m)$ can be expanded around $m=0$ as~\cite{aron2020landau}
\begin{align} \label{eq:exppotcw}
V(m) =&\frac{1}{2} \left(1-\frac{T_c}{T_1} \right) m^2 +B |m|^{2+\nu}+O(m^4) \,,
\end{align}
where
\begin{align}
  B= \frac{\gamma_{2} (k_B T_c)^{\nu+1}}{\gamma_1 k_B T_1 (2+\nu)} \left(\frac{T_2}{T_1}-1 \right) \,,
\end{align}
and $T_c=J/k_B$ is the critical temperature of the equilibrium Curie--Weiss model. Notably, the expansion includes the nonanalytic term proportional to $|m|^{2+\nu}$. This term appears only out of equilibrium, as it vanishes for $T_1=T_2$. Physically, it is related to different low-energy behavior of the spectral densities of the baths.

Let us note that in our paper we use the exact formula for the quasipotential [Eq.~\eqref{eq:quasipotform}], while in Ref.~\cite{aron2020landau} it was calculated using the Fokker-Planck equation. It is known that the latter approach incorrectly evaluates the quasipotential away from its minima and saddle points~\cite{FalascoReview, hanggi1988bistability, gaveau1997master, vellela2009stochastic, gopal2022large, hanggi1982stochastic}. However, we verified that the methods agree with respect to both the quadratic and nonanalytic terms of the expansion. The discrepancy appears only for higher-order terms, which are inconsequential for the later results.

\subsection{Critical behavior of the system}
\begin{figure}
	\centering
	\includegraphics[width=0.9\linewidth]{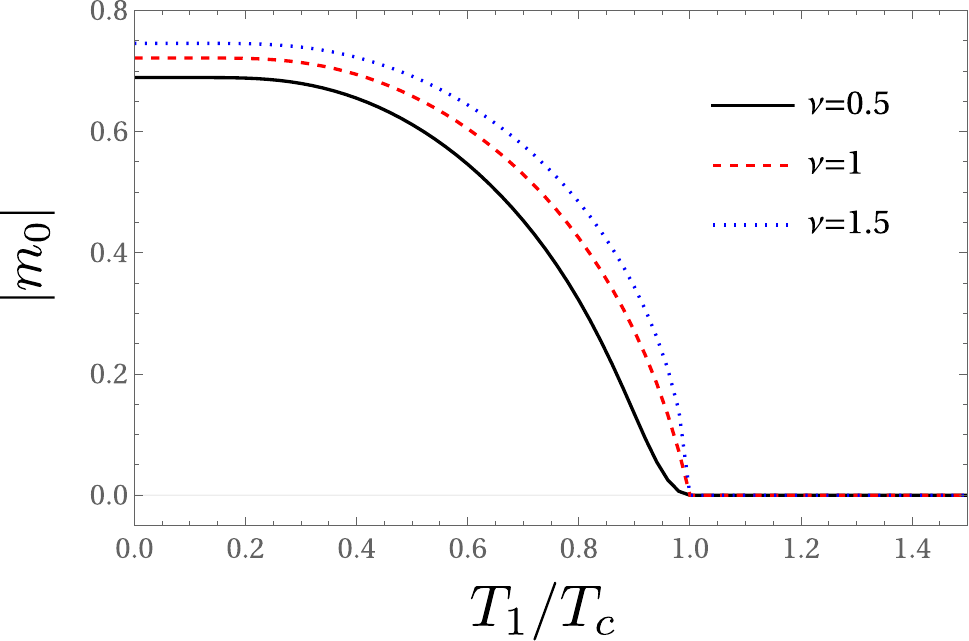}
	\caption{The normalized stationary magnetization as a function of the bath temperature $T_1$ for different values of the parameter $\nu$. Parameters: $\alpha_1=1$, $T_2=1.25T_c$, $\gamma_2=\gamma_1 J^{-\nu}$.}
	\label{fig:magn}
\end{figure}

Let us now briefly summarize the results of Ref.~\cite{aron2020landau} concerning the critical behavior of the system considered. We focus on the case of $\nu \in (0,2)$, where the nonanalytic term dominates over the quartic term. We also take $T_2>T_1$, so that the system exhibits a continuous phase transition (otherwise it is discontinuous). We denote the position of the global minimum of $V(m)$, which corresponds to normalized (divided by $N$) stationary magnetization of the system, as $m_0$:
\begin{align}
m_0=\argmin_{m \in [-1,1]} V(m) \,.
\end{align}
As implied by Eq.~\eqref{eq:exppotcw}, the magnetic ordering of the system is determined only by the lowest temperature $T_1$. This is because for $\alpha_1<\alpha_2$ the bath $1$ is more strongly coupled to the low-energy excitations of the system. For $T_1 \geq T_c$, the system is in the paramagnetic state, with a single minimum of $V(m)$ at $m_0=0$. In contrast, for $T_1<T_c$, the system has two degenerate minima at $m_0=\pm|m_0|$, which correspond to the opposite magnetization states. Thus, at $T_1=T_c$, the system undergoes a symmetry-breaking phase transition. For $T_1$ smaller but close to $T_c$, the stationary magnetization exhibits a power-law scaling
\begin{align} \label{eq:magncritexp}
|m_0| \propto \left( \frac{T_c-T_1}{T_2-T_1} \right)^{1/\nu} \,.
\end{align}
Thus, the nonanalytic term of the Landau functional determines the critical exponent of the phase transition $\hat{\beta}=1/\nu$, which is in general larger than the equilibrium mean field critical exponent $\hat{\beta}_\text{MF}=1/2$. In Fig.~\ref{fig:magn} we present the exact dependence of the stationary magnetization $m_0$ on temperature $T_1$. As shown, the parameter $\nu$ affects the magnetization behavior near the critical point $T_1=T_c$, in agreement with Eq.~\eqref{eq:magncritexp}, as well as far from the critical point. We can also see that for $T_1 \rightarrow 0$ the normalized magnetization $m_0$ does not tend to $1$ due to the disordering effect of the bath 2.

 %
%

In analogy to the molecular zipper, the nonanalytic term also determines the order of the phase transition. Since the system is out of equilibrium, the order is no longer defined by the behavior of free energy. Still, we may define the order of phase transition by the behavior of the order parameter: the transition is of $j$th order if the $(j-1)$th derivative of $|m_0|$ over $T_1$ is the lowest, which is discontinuous at the phase transition point $T_1=T_c$. Thus, the transition is of the $j$th order when
\begin{align}
    \frac{1}{j-1} \leq \nu <\frac{1}{j-2} \,,
\end{align}
which is analogous to Eq.~\eqref{eq:ordzipp} for the molecular zipper. As for the zipper, $j \geq 2$, since the considered phase transition is continuous. We note that Ref.~\cite{aron2020landau} used another convention, which distinguished only second- and third-order phase transitions.

\subsection{Magnetization fluctuations}
Let us now analyze how the nonanalytic term of the Landau functional influences the scaling of fluctuations at the phase transition point $T_1=T_c$, $h=0$. First, we consider the magnetization moments defined as
\begin{align}
\langle M^{k} \rangle \equiv \sum_{M} p_M^\text{st} M^k \,.
\end{align}
They can be calculated using stationary probabilities given by Eq.~\eqref{eq:stprob}. In numerical calculations, this allows one to study systems consisting of millions of spins. As implied by Eq.~\eqref{eq:encwm}, for $h=0$, the second moment of the magnetization is related to the average energy as
\begin{align}
\langle E \rangle=-\frac{J}{N} \langle M^2 \rangle \,.
\end{align}
Going to the continuous limit, we can approximate the magnetization moments as
\begin{align} \label{eq:magnmom-int}
\langle M^k \rangle \approx N^k \frac{\int_{-\infty}^\infty m^k e^{-N V(m)} dm}{\int_{-\infty}^\infty e^{-N V(m)} dm} \,.
\end{align}
Since the behavior of the exponential function $\exp[-N V(m)]$ is dominated by the leading order of $m$, we can replace $V(m)$ with the nonanalytic term $B |m|^{2+\nu}$. The solution yields
\begin{align} \label{eq:magnmom}
\langle M^k \rangle \approx \frac{1+(-1)^k}{2} N^{\frac{k(1+\nu)}{2+\nu}}  B^{-\frac{k}{2+\nu}} \frac{\Gamma \left(\frac{1+k}{2+\nu} \right)}{\Gamma \left(\frac{1}{2+\nu} \right)} \,.
\end{align}
One can observe that in the case considered, the odd moments of $M$ vanish because of the $\mathbb{Z}_2$ symmetry of the model (that is, the symmetry with respect to the magnetization reversal $M \rightarrow -M$). We also note that when the nonanalytic term is absent or the quartic term of the Landau potential dominates (i.e., $\nu \geq 2$), the even moments obey a universal scaling relation $\langle M^k \rangle \propto N^{\frac{3k}{4}}$.

 %
\begin{figure}
	\centering
	\includegraphics[width=0.9\linewidth]{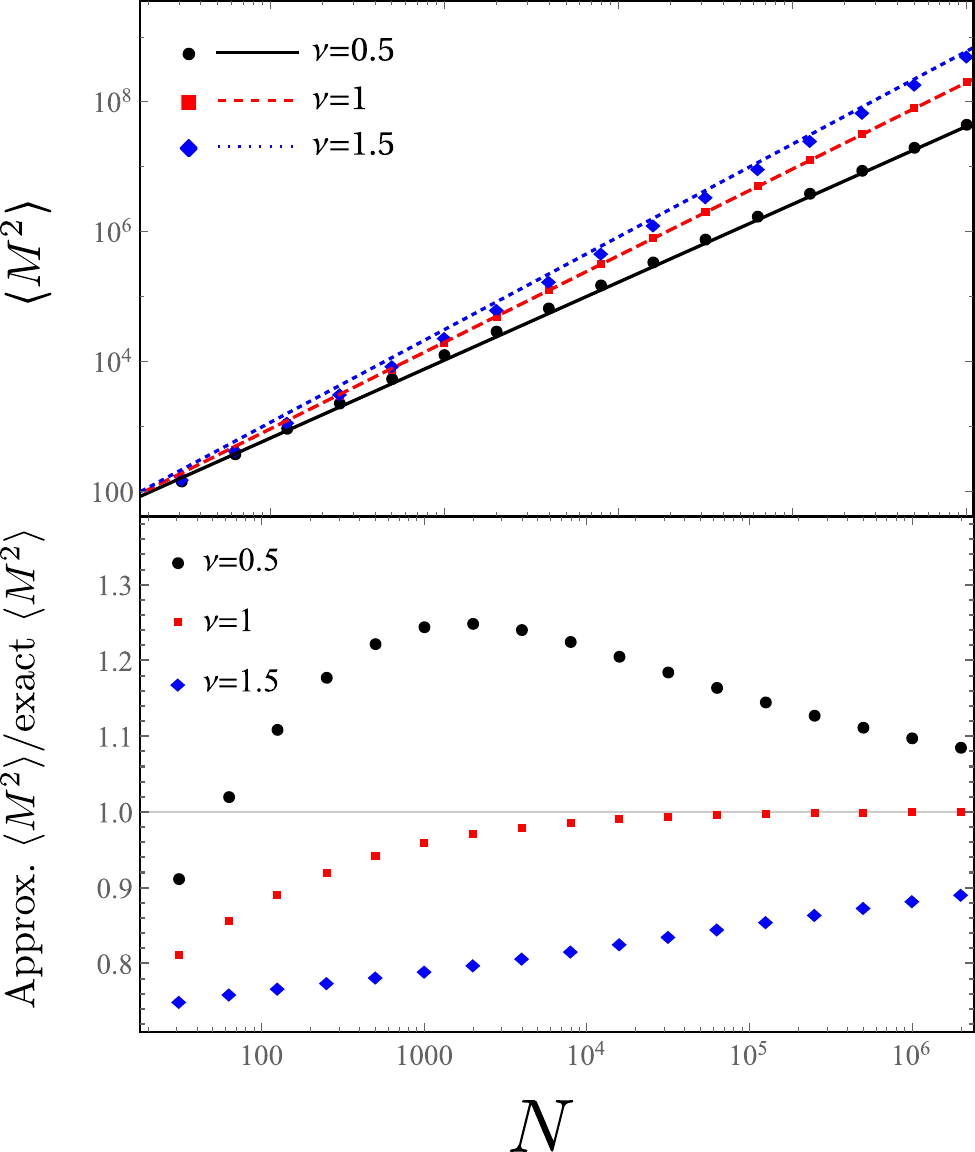}
	\caption{(a) The second moment of magnetization $\langle M^2 \rangle$ at the critical point $T_1=T_c$ as a function of the system size $N$, plotted on the log-log scale, calculated exactly (dots) and using the approximate formula~\eqref{eq:magnmom} (lines). (b) Ratio of approximate and exact results for $\langle M^2 \rangle$, plotted on the log-linear scale. Parameters as in Fig.~\ref{fig:magn}.}
	\label{fig:secondmom}
\end{figure}

The exemplary finite-size scaling of the second moment of magnetization $\langle M^2 \rangle$ (dots), compared to the above approximation (lines), is plotted in Fig.~\ref{fig:secondmom}~(a). As shown, Eq.~\eqref{eq:magnmom} correctly describes the character of the power-law scaling. This confirms the effect of the nonanalytic term of the Landau functional on the finite-size scaling of fluctuations of magnetization. However, as shown in Fig.~\ref{fig:secondmom}~(b), for $\nu=0.5$ or $\nu=1.5$ the difference between the approximate and exact results is of the order of $10\%$ even for systems consisting of millions of spins. This may be related to higher-order terms of the Landau functional that are neglected in our approximation.

\subsection{Binder cumulant}
To further illustrate the effect of the nonanalytic term, let us analyze the parameter known as the \textit{Binder cumulant}. It is defined as~\cite{binder1981finite,binder1981critical}
\begin{align}
U \equiv 1-\frac{\langle M^4 \rangle}{3\langle M^2 \rangle^2} \,.
\end{align}
This parameter quantifies the kurtosis of the magnetization probability distribution. Its peculiar feature is that, with increasing system size, it asymptotically converges to a certain finite value. Using Eq.~\eqref{eq:magnmom}, it can be approximated as
%
\begin{figure}
	\centering
	\includegraphics[width=0.9\linewidth]{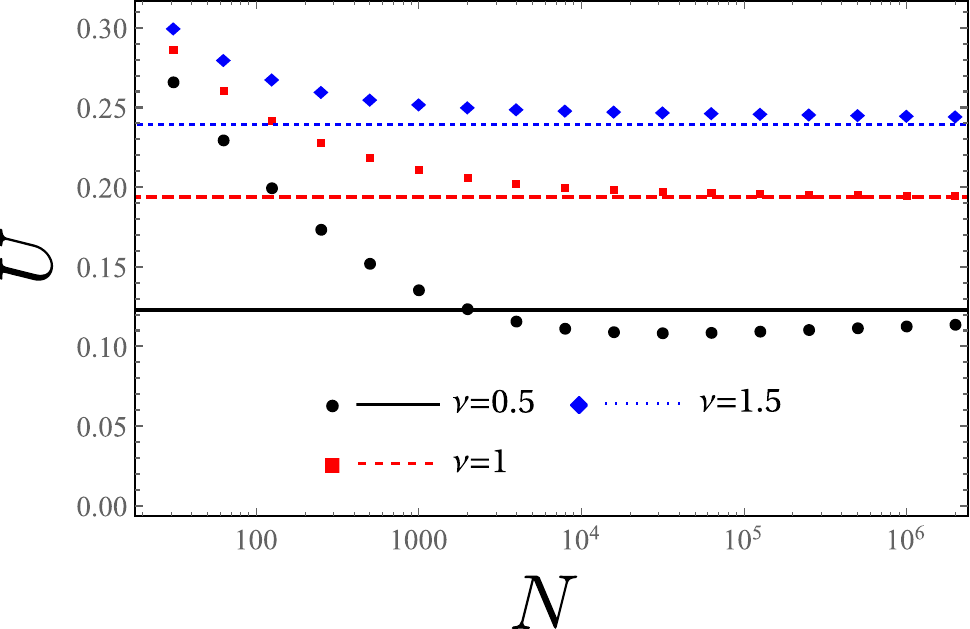}
	\caption{The Binder cumulant at the critical point $T_1=T_c$ as a function of the system size $N$, plotted on the log-linear scale, calculated exactly (dots) and using the approximate formula~\eqref{eq:binderapp} (lines). Parameters as in Fig.~\ref{fig:magn}.}
	\label{fig:binder}
\end{figure}
%
\begin{align} \label{eq:binderapp}
U \approx1-\frac{\Gamma \left(\frac{1}{\nu +2}\right) \Gamma \left(\frac{5}{\nu +2}\right)}{3 \Gamma \left(\frac{3}{\nu +2}\right)^2} \,,
\end{align}
which characteristically depends only on the parameter $\nu$, and not on the system size nor the amplitude of the nonanalytic term. The exact scaling of the Binder cumulant (dots) compared with the above approximation (lines) is presented in Fig.~\ref{fig:binder}. As shown, Eq.~\eqref{eq:binderapp} well approximates the asymptotic value of the Binder cumulant. However, similarly to the behavior of $\langle M^2 \rangle$, the convergence to the asymptotic value is quite slow for $\nu=0.5$ or $\nu=1.5$. Additionally, the Binder cumulant for $\nu=0.5$ exhibits a characteristically nonmonotonic behavior, going first below the asymptotic value and then approaching it from below. 

\subsection{Magnetic susceptibility} \label{subsec:susc}
Finally, let us analyze the finite-size scaling of the magnetic susceptibility, i.e., the response of average magnetization to the magnetic field, evaluated at the phase transition point. It is defined as
\begin{align}
\chi \equiv \left( \frac{\partial \langle M \rangle}{\partial h} \right)_{h=0} \,,
\end{align}
and can be calculated as
\begin{align} \label{eq:susc-calc}
\chi=\sum_M M \left( \frac{\partial p_M^\text{st}}{\partial h} \right)_{h=0} \,.
\end{align}
For finite system sizes, the derivatives of probabilities $p_M^\text{st}$ can be obtained as~\cite{ptaszynski2024critical}
\begin{align}
 \frac{\partial \boldsymbol{p}^\text{st}}{\partial h} =-\mathbb{W}^D \frac{\partial \mathbb{W}}{\partial h}\boldsymbol{p}^\text{st} \,,
\end{align}
where the probability vector $\boldsymbol{p}^\text{st}$ and the rate matrix $\mathbb{W}$ were defined below Eq.~\eqref{eq:masteqmat}, and $\mathbb{W}^D$ is the Drazin inverse of the rate matrix (see Refs.~\cite{crook2018drazin,landi2023current} for its definition and properties). For an alternative method to determine the static response of the stationary state, see Refs.~\cite{aslyamov2024nonequilibrium,aslyamov2024general}.

%
\begin{figure}
	\centering
	\includegraphics[width=0.9\linewidth]{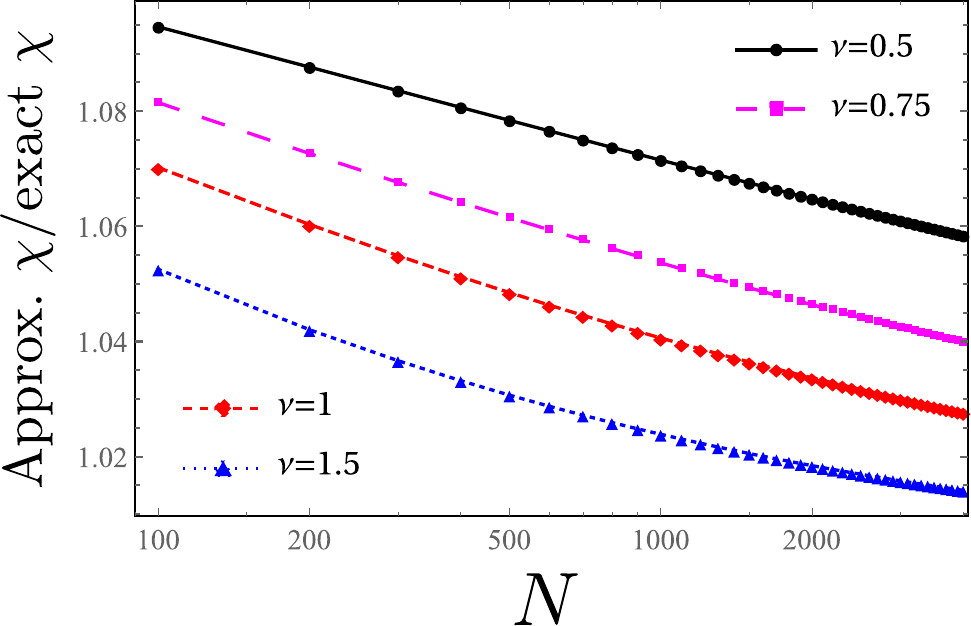}
	\caption{The ratio of the magnetic susceptibility at the critical point $T_1=T_c$ approximated using Eq.~\eqref{eq:susc2mom} to the exact susceptibility given by Eq.~\eqref{eq:susc-calc}, plotted as a function of the system size $N$. The results are represented by dots and plotted on the log-linear scale. The lines are added for eye guidance. Parameters as in Fig.~\ref{fig:magn}.}
	\label{fig:ratiosusc}
\end{figure}
%
Let us now go to the continuous limit. We now denote the nonequilibrium quasipotential as $V(m,h)$, explicitly indicating its dependence on the magnetic field. The magnetic susceptibility can be approximated using Eq.~\eqref{eq:magnmom} as
\begin{align} \label{eq:suscept-exp} \nonumber
\chi & \approx N  \left[\frac{\partial }{\partial h} 
 \frac{\int_{-\infty}^\infty m e^{-N V(m,h)} dm}{\int_{-\infty}^\infty e^{-N V(m,h)} dm} \right]_{h=0}  \\ \nonumber
& =-N^2 \frac{\int_{-\infty}^\infty m e^{-N V(m,0)} \partial_{h=0} V(m,h) dm}{\int_{-\infty}^\infty e^{-N V(m,0)} dm} \\
&-N
 \frac{\int_{-\infty}^\infty m e^{-N V(m,0)} dm}{\left[\int_{-\infty}^\infty e^{-N V(m,0)} dm \right]^2} \left[\frac{\partial }{\partial h} 
 \int_{-\infty}^\infty e^{-N V(m,h)} dm \right]_{h=0} \,,
\end{align}
where we denoted
\begin{align}
\partial_{h=0} V(m,h)=\left[ \frac{\partial{V(m,h)}}{\partial h} \right]_{h=0} \,.
\end{align}
The second term in the last expression in Eq.~\eqref{eq:suscept-exp} vanishes because of $\int_{-\infty}^\infty m e^{-N V(m,0)} dm=0$. The derivative $\partial_{h=0} V(m,h)$ can be expanded at $T_1=T_c$ as 
\begin{align} \label{eq:dervh}
\partial_{h=0} V(m,h)=-\frac{m}{k_B T_c} + O(|m|^{2+\nu}) \,.
\end{align}
Inserting this into Eq.~\eqref{eq:suscept-exp}, subtracting the higher orders in $m$, which can be neglected for large $N$, and comparing the resulting expression with Eq.~\eqref{eq:magnmom-int}, we find that the magnetic susceptibility can be related to the magnetization second moment as
\begin{align} \label{eq:susc2mom}
\chi \approx \frac{\langle M^2 \rangle}{k_B T_c} \,.
\end{align}
Thus, even though the system is out of equilibrium, the magnetic susceptibility and the second moment of magnetization are related via an equilibrium-like formula. However, we note that this holds only asymptotically for large $N$ rather than exactly, as in the equilibrium case. 

Let us now explore the validity of the formula~\eqref{eq:susc2mom} for a finite $N$. This is presented in Fig.~\ref{fig:ratiosusc}. Due to the necessity of calculating $\mathbb{W}^D$, our study is now limited to smaller systems, with $N$ of the order $10^3$. As shown, for the cases presented, this approximation overestimates the magnetic susceptibility only by a few percent, with the agreement improving with $N$. Thus, we may infer that Eq.~\eqref{eq:susc2mom} becomes asymptotically exact in the thermodynamic limit. Furthermore, the approximation works better for higher values of $\nu$, since the higher orders of expansion~\eqref{eq:dervh} are then less important. Therefore, for large $N$, the magnetic susceptibility obeys the same scaling as the second moment of magnetization. As shown before, the latter can be approximated using Eq.~\eqref{eq:magnmom}, and thus depends on the parameter $\nu$, which characterizes the exponent of the nonanalytic term of the Landau functional.

\section{Conclusions} \label{sec:concl}
We have shown that the nonanalytic terms of the Landau functional can determine the finite-size scaling of fluctuations and response functions at the continuous phase transition points. This was demonstrated on the equilibrium molecular zipper model and the nonequilibrium version of the Curie--Weiss model. In particular, using large deviation theory, we derived approximate analytic power-law formulas describing the finite-size scaling, whose scaling exponents are determined by the nonanalytic terms of the Landau functional. For the molecular zipper model, we observed a very good agreement between those formulas and the exact results even for relatively small systems. For the nonequilibrium Curie--Weiss model, these formulas also accurately describe the scaling exponents. However, the approximate and exact values of the fluctuations and responses differ slightly, but notably, even for relatively large systems. This might be ascribed to the effect of the higher-order terms of the Landau functional.

The demonstrated explicit relation between the power-law scaling exponents of fluctuations or responses and the nonanalytic terms of the Landau functional may allow determining the presence and form of these nonanalytic terms -- that can shape the critical behavior of the system in the thermodynamic limit -- through measurements or simulations of finite-size systems. However, we emphasize that our study focused on mean field models, for which the Landau theory is exact. The open question is whether our conclusions can be generalized to finite-dimensional systems. The latter may encompass equilibrium systems with soft modes, whose Landau functional include nonanalytic terms~\cite{belitz2005how}, or nonequilibrium spin lattices, analogous to the model analyzed in Sec.~\ref{sec:cw}. On the one hand, an affirmative answer to that question is supported by the fact that the behavior of systems near criticality can be effectively captured using the renormalization group techniques. These methods build upon Landau theory, refining it to account for spatial correlations~\cite{wilson1974renormalization}. On the other hand, in nonequilibrium spin lattices, it has been questioned whether the effect of nonanalytic behavior of spectral densities--responsible for the formation of nonanalytic Landau functionals--might be suppressed by the discrete nature of the effective magnetic field acting on spins (which is determined by the discrete configuration of a few neighboring spins)~\cite{aron2020landau}. However, a later study provided some evidence that nonanalytic Landau functionals may in fact play a role in shaping magnetization fluctuations also in finite-dimensional lattices~\cite{aron2021nonanalytic}.

\section*{Data availability}
Wolfram Mathematica notebooks used to obtain the numerical results and the data used in the figures are available at the following DOI: 10.5281/zenodo.14826139

\begin{acknowledgments}

K.P.\ acknowledges the financial support of the National Science Centre, Poland, under the project No.\ 2023/51/D/ST3/01203.

\end{acknowledgments}

\bibliography{bibliography}	
	
\end{document}